\documentclass[12pt,onecolumn]{article}
\usepackage{lipsum}

\usepackage{times}
\usepackage{xr}
\usepackage{graphicx}
\usepackage{bm}
\usepackage{amsmath}
\usepackage{amssymb}
\usepackage[dvipsnames]{xcolor}
\usepackage{hyperref}
\usepackage{siunitx,xfrac}
\usepackage{xcolor}

\usepackage{etoolbox}
\usepackage{nomencl}
\makenomenclature
\newcommand{\diff}{\mathop{}\!\mathrm{d}}
\usepackage[left=0.5in, right=0.5in, top=0.5in, bottom=1in]{geometry}
\usepackage{cite}

\title{Building granular structures with elasto-active systems} 

\author{
Yuchen Xi$^{1}$, Tom Marzin$^{1}$, P.-T. Brun$^{1,2}$.
\\
\small{$^{1}$Department of Chemical and Biological Engineering,}
\small{Princeton University, Princeton, NJ 08540, USA}\\
\small{$^{2}$Department of Chemical Engineering, Soft Matter, Rheology and Technology(SMaRT), KU Leuven, Leuven, Belgium}
}

\date{}

\begin{document}
\maketitle

\begin{abstract}
Natural active systems routinely reshape and reorganize their environments through sustained local interactions. Examples of decentralized collective construction are common in nature, e.g., many insects achieve large-scale constructions through indirect communication. While synthetic realizations of self-organization exist, they typically rely on rigid agents that require some kind of sensors and direct programming to achieve their function. Understanding how soft, deformable active matter navigates and remodels crowded landscapes remains an open challenge. Here we show that connecting rigid microbots to elastic beams yields elasto-active structures that can restructure and adapt to heterogeneous surroundings. We investigate the dynamics of these agents in environments with varying granular densities, rationalizing how they can aggregate or carve the medium through gentle interactions. At low density, the system compacts dispersed obstacles into clusters, a process modeled by a modified Smoluchowski coagulation theory. At high density, our agents carve voids whose size is predicted by a force-limited argument. These results establish a framework for understanding how activity, elasticity, and deformability can influence active navigation and environmental reconfiguration in granular media.

\end{abstract}

\section*{Significance}

Soft organisms, from worms burrowing through soil to flagella stirring in complex and crowded media, reshape their environments using bodies that are inherently flexible and adaptive. Drawing inspiration from these systems, we demonstrate that simple synthetic structures, composed of active particles connected by an elastic beam, can autonomously reorganize granular materials through purely physical interactions. This emergent behavior shows how softness alone can confer adaptive, environment-modifying abilities to minimal agents. Our findings reveal a route toward designing decentralized soft systems capable of collective construction, with potential implications for manufacturing, robotics, and sustainable material processing.

\section*{Introduction}
Collectives of active motile units, from bacterial swarms to flocks of birds, can exhibit autonomous, coherent, and large-scale dynamics due to purely local interactions\cite{marchetti2013hydrodynamics}. In these systems, energy-consuming units organize themselves into spatiotemporal patterns that far exceed the complexity of any single constituent. A broad range of synthetic active matter systems, such as self-propelled colloids\cite{Bricard_colloidal_rollers}, vibrated granular particles\cite{Deseigne_disk}, and centimeter-scale robots\cite{hexbug,xia_celia_2024,baconnier2022selective,ben_zion_morphological_2023,dauchot2019dynamics,giomi_swarming_2013,VidalesHernndez2024}, have been developed to explore these dynamics. However, most of these artificial systems are typically composed of rigid units whose interactions are dominated by hard steric collisions or short-range forces, constraining their responsiveness and adaptability in heterogeneous or evolving environments.

In contrast, many living active systems are intrinsically soft and deformable. From cilia that rhythmically transport fluids to worms and slime molds that contort their bodies to navigate tortuous terrains \cite{Kanale_Ling_Guo_Fürthauer_Kanso_2022,Pan2025, locomotionpolymerlike}, deformability is not merely a feature but a functional advantage. It enables organisms to traverse, adapt to, and physically reshape their surroundings, capabilities beyond the reach of rigid-body systems. Motivated by these natural strategies, recent efforts have begun to explore flexible active structures in the laboratory. These include self-oscillating gels that autonomously bend and contract through internal chemical reactions\cite{Yoshida2010}, magnetically driven filaments that sweep and transport particles by modulating their curvature\cite{Cebers2016,Sun2022}, and elasto-active structures that harness morphological computation to achieve predictive motion and complex environment navigation\cite{zheng_self-oscillation_2023,Xi2024elastoactive,vatin2025translocationactivepolymerlikeworms,Son2025}.

Another key feature of active matter systems in nature is that they rarely operate in empty environments. Instead, they inhabit heterogeneous, reconfigurable environments, such as dense cellular matrices, porous soils, and fibrous tissues\cite{activematter_review,Needleman2017, Hallatschek2023,Jin2024}. In such settings, locomotion is linked to environmental restructuring, where their motion modifies local terrain, which, in turn, impacts the system's dynamics. Unraveling the mechanisms by which soft, deformable structures move through and reshape densely structured environments remains an open question. Such interactions can induce structure formation\cite{schunter_elastogranular}, and aggregation phenomenon\cite{sinaasappel2025collectingparticlesconfinedspaces} that echo classical coagulation processes. While Smoluchowski kinetics has originally been used to describe clustering in passive systems \cite{smoluchowski1918versuch,friedlander2000smoke}, such as colloidal particles \cite{smoluchowski1918versuch}, its relevance to active to shape-changing agents navigating dynamic terrains has only recently been discovered~\cite{Xi2024elastoactive,sinaasappel2025collectingparticlesconfinedspaces}, and our knowledge of aggregation in this context remains in its infancy. 

\begin{figure*}[!ht]
\begin{center}
    \includegraphics[width=1\textwidth]{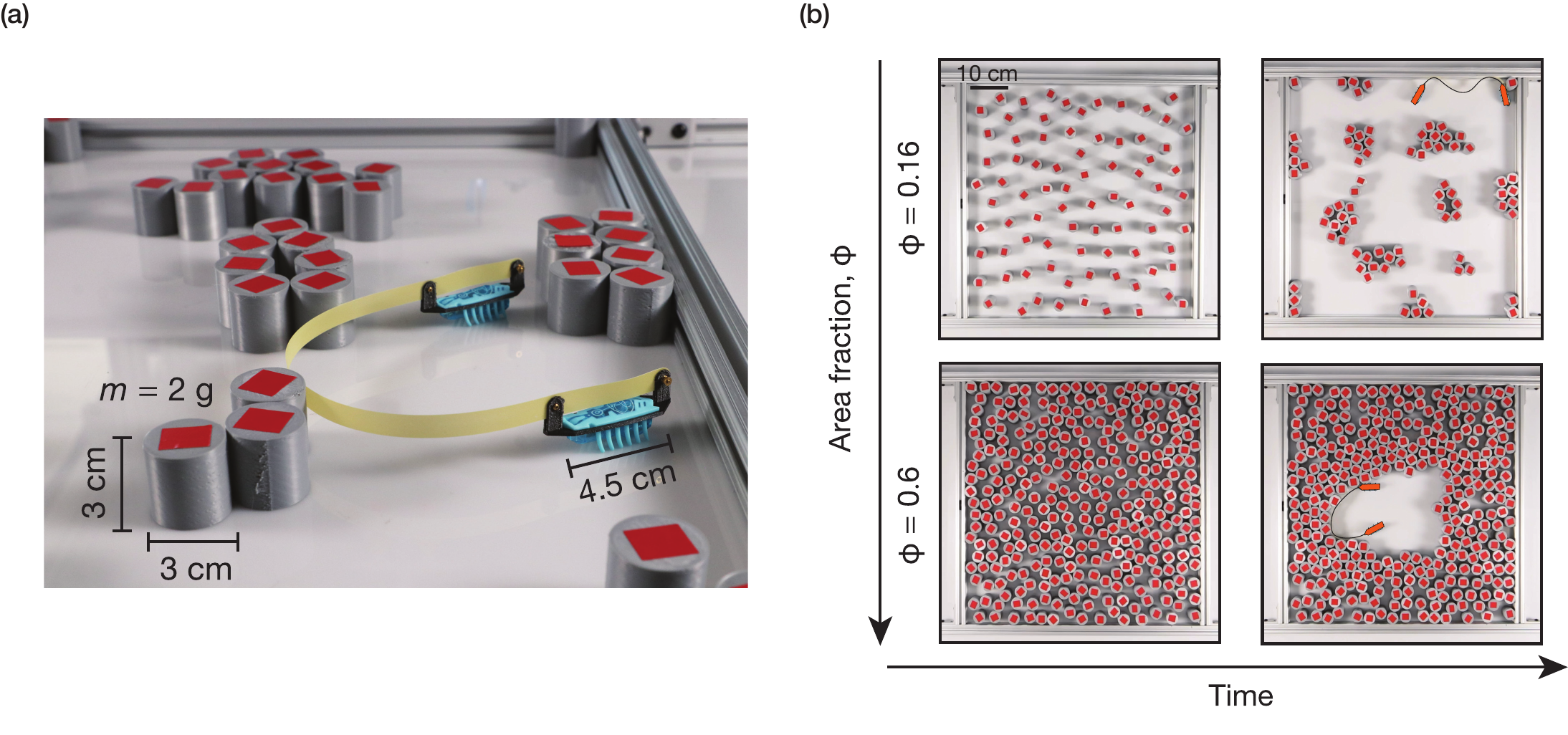}
    \caption{\textbf{Elasto-active structure interacts with passive granular medium} (a) Experimental snapshot showing the details of the elasto-active structure and the cylindrical particles. The structure consists of two active microbots connected via an elastic beam. The gray cylindrical particles are 3D printed, and the red stickers are used for tracking purposes. (b) At low solid fraction (density), the elasto-active structure pushes obstacles into clusters. At a high solid fraction (density), the structure carves a void within the medium.}
    \label{fig1}
    \end{center}
\end{figure*}

Fig. \ref{fig1}(a) illustrates our approach: we explore the dynamics of an elastic active structure that reshapes and organizes a reconfigurable granular environment through sustained interactions. Our structures, or bucklebots \cite{Xi2024elastoactive}, consist of two centimeter-scaled active microbots connected by an elastic beam. These bucklebots adopt a V-shape and travel across space, demonstrating distinct behaviors depending on the granular densities (see Fig. \ref{fig1}(b)). At relatively low solid fractions, the buckelbot compacts an initially dispersed field of obstacles into discrete clusters. At high densities, by contrast, our robotic system carves a void enclosed by a dense, mechanically reinforced boundary. To interpret these dynamics, we integrate experiments with two complementary theoretical approaches: a kinetic model inspired by Smoluchowski coagulation theory\cite{smoluchowski1918versuch}, and a geometrical framework that captures the constraints imposed by obstacle crowding and force transmission. Together, these results reveal principles by which soft active agents can navigate, interact, and restructure within complex environments.

\section*{Results}

\begin{figure*}[!ht]
\begin{center}
    \includegraphics[width=1\textwidth]{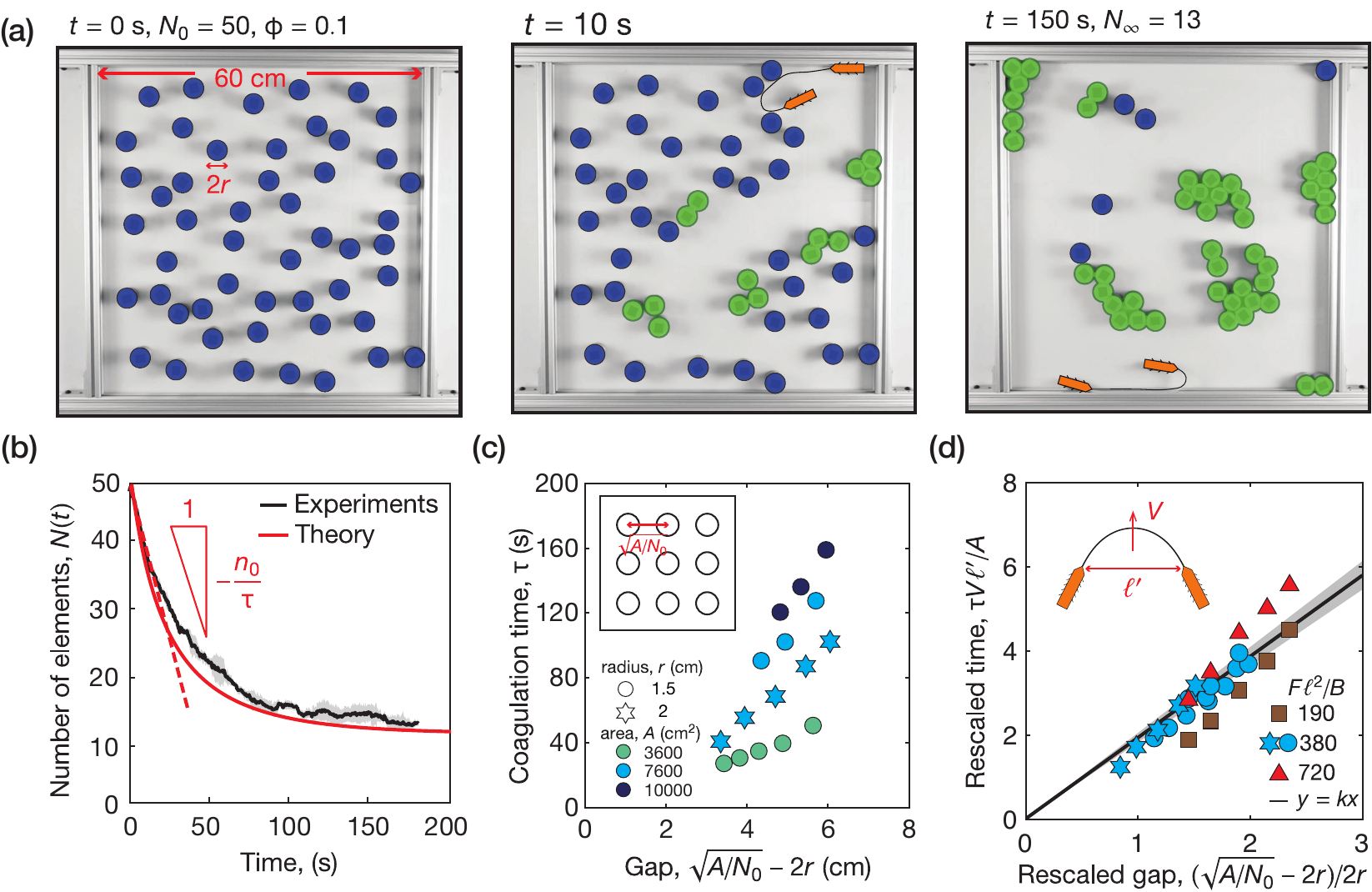}
    \caption{\textbf{Coagulation dynamics in low-packed medium} (a) Snapshots of the experiment with a density of obstacles $\phi \simeq 0.1$ and with a bucklebot with parameter$F\ell^2/B \simeq 380$. The dark blue circles denote isolated obstacles, and the green boundaries denote formed clusters. (b) The number of elements $N(t)$  (single obstacles and clusters) is plotted against time. The shaded area denotes the standard deviation within three trials. The dashed red line is the initial slope when taking the limit $n(t)\simeq n_0$ (See SI Eqn. \ref{soln_coagulation1}). The solid red line is the theory combining coagulation and fragmentation dynamics. (c) Plot of the coagulation timescale $\tau$ versus the effective gap parameter $\sqrt{A/N_0}-2r$. The markers are color-coded by area, while different shapes correspond to different types of obstacles. Error bars denote the standard deviation within 3 trials. Inset: definition of the effective gap parameter. (d) Rescaled time $\tau V\ell' /A$ versus rescaled gap. The markers are color-coded by the rescaled forces of the bucklebots. The black line is a linear fit of $y=kx$ with $k=1.93\pm0.06$, where the shaded band indicates the 95\% confidence interval of the fit. Inset: a sketch of the bucklebot showing its equilibrium width $\ell'$ and velocity $V$.}
    \label{fig2}
    \end{center}
\end{figure*}

\begin{figure*}[!ht]
\begin{center}
    \includegraphics[width=0.8\textwidth]{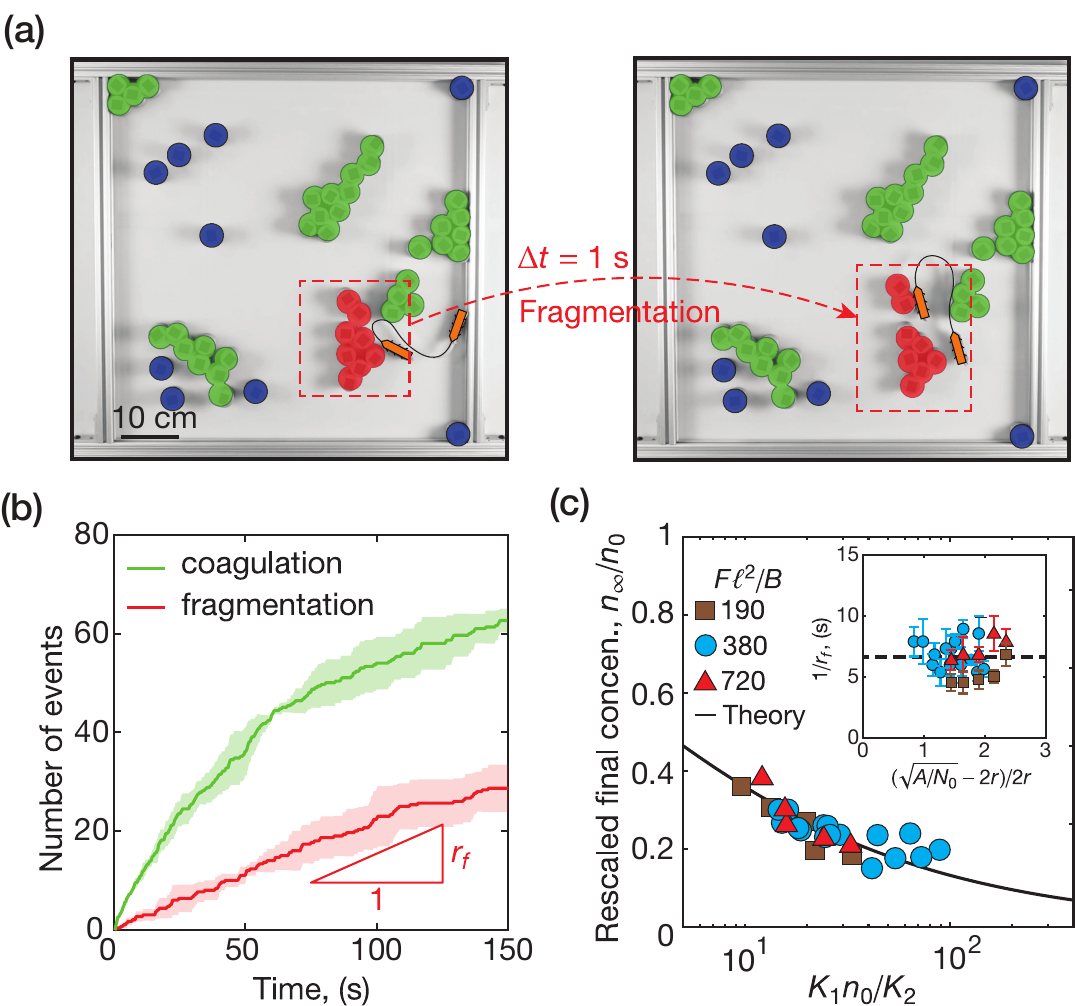}
    \caption{\textbf{Fragmentation dynamics} (a) Consecutive snapshots showing a bucklebot fragments a cluster into two separate clusters, highlighted by the red boundaries. (b) The number of coagulation(green) and fragmentation(red) events plotted against time. Each shaded area denotes the standard deviation within five trials. The slope of the fragmentation line gives the global fragmentation rate $r_f$. (c) Rescaled final element concentration $n_\infty/n_0$ versus the evolution parameter $K_1n_0/K_2$. The markers are color-coded by the rescaled forces of the bucklebots. The black line represents the prediction from the theory. The inset shows the value of the fragmentation time $1/r_f$ across all experiments. The black dashed line represents the mean.}
    \label{fig3}
    \end{center}
\end{figure*}

Figures \ref{fig2} and \ref{fig3} summarize the main results pertaining to bucklebots interacting with obstacles they can displace in a moderately packed granular medium (the solid fraction ranges from $\phi \simeq 0.1$ to $\phi \simeq 0.2$). In particular, Fig. \ref{fig2}(a) shows a bucklebot confined in a square space with area $A=3600$ cm\textsuperscript{2} , containing $N_0=50$ dispersed cylindrical obstacles with radius $r=1.5$ cm and mass $m=2$ g. The bucklebot\cite{Xi2024elastoactive}, is bent  as the rescaled force $F\ell^2/B\simeq380\gg 1$, where $F \simeq 20$ mN is the active force of the microbot, $B$ is the bending stiffness, and $\ell$ is the length of the connecting beam. It pushes these light obstacles and assembles them into clusters (See Movie S1). In the figure,  green marks the clusters that form, while blue denotes single obstacles. In Fig. \ref{fig2}(b), we report the value of  $N(t)$ within the space, defined as the number of clusters of any size. We find that $N(t)$ decays rapidly and then saturates to a steady-state value two minutes after the experiment started. The number of elements at equilibrium $N_\infty=13$ is such that a storing ratio  $(N_0-N_\infty)/N_0=0.74\%$ of obstacles are stored into clusters.  In comparison, microbots alone (that is, without any elastic connection) only store 38\% of obstacles into clusters~\cite{Xi2024elastoactive}.

We model the short-term dynamics of our problem with a Smoluchowski-like coagulation equation \cite{smoluchowski,fair1964mathematical} (SI Section I), derived from a balance of elements' concentration $n(t)=N(t)/A$. From experiments, we extract the coagulation timescale $\tau$ from the slope at the onset, where $n(t)\simeq n_0$. In Fig. \ref{fig2}(c), we show the variation of this timescale when the area and radius of the obstacle are varied. We plot $\tau$ against an effective spacing parameter $\sqrt{A/N_0}-2r$, which represents the mean initial gap between two obstacles, or equivalently, the average distance a bucklebot travels before encountering an obstacle. The coagulation timescale increases with the gap for each parameter set, indicating that larger initial gaps lead to a slower coagulation process. In Fig. \ref{fig2}(d), we recast our experimental data in dimensionless form using $2r$ as the length gauge and $A/V\ell'$ as the time gauge. This timescale captures the bucklebot's efficiency to cover free space, as $V\simeq15$ cm/s is its velocity and $\ell'$ is the equilibrium width defined as the distance between two ends of the beam when the bucklebot travels freely in its equilibrium configuration. Both quantities depend on the rescaled force parameter $F\ell^2/B$,\cite{Xi2024elastoactive}. Additional data where we vary the value of $F\ell^2/B$ by modifying the beam's properties are shown on the plot. The data collapse to a single master curve, in particular confirming the relevance of the initial distribution of the obstacles $\sqrt{A/N_0}-2r$ on the speed of coagulation. As obstacles are further apart, collisions between bucklebots and obstacles become less frequent, resulting in slower cluster formation and longer coagulation time. Furthermore, the data show a linear relationship between space and time with a slope $k=1.93\pm0.06$. Such a simple relationship indicates that the coagulation dynamics are strongly determined by the rate at which the bucklebot travels across space. More complex effects only come into play at a later time, as explained next. 

Having understood the short-term coagulation dynamics of the system, we move to predicting the long-term equilibrium state. In Fig. \ref{fig3}(a), consecutive experimental snapshots show that, while the bucklebots initially push obstacles into clusters, they can also fragment clusters. In Fig. \ref{fig3}(b), we show the number of coagulation and fragmentation events over time. The frequency of coagulation events is initially higher than that of fragmentation events. Yet, both coagulation and fragmentation then proceed at a nearly constant rate. After about 100 seconds, the two curves exhibit comparable slopes, indicating a balance between the two processes and the emergence of a dynamic steady state. To quantitatively account for this phenomenon, we add a fragmentation term to the Smoluchowski-like equation (See SI Section II). 
\begin{equation}
    \frac{\diff{n(t)}}{\diff{t}}=\underbrace{-\frac{1}{2}K_1n(t)^2}_\text{coagulation} + \underbrace{K_2(n_0-n(t))}_\text{fragmentation}, \ n(0)=n_0
\end{equation}
where $n(t)=N(t)/A$ is the concentration of clusters of $N$ particles. This model now has two terms that account for coagulation and fragmentation, with $K_1$ and $K_2$ being their respective rate constants (note their units differ). The value of $K_1$ is given by the initial coagulation timescale as $K_1=2/(n_0\tau)$ (See SI Section II). $K_2$ captures the probability of cluster fragmentation, assumed to be constant for simplicity. This choice is motivated by the data reported in Fig. \ref{fig3}(b), where we extract $r_f$, the fragmentation rate. We translate this rate into our constant $K_2=r_f/(n_0 A)$, which can be interpreted as the frequency at which a single particle participates in a fragmentation event  (See SI Section II). Guided by this model, we recast our experimental data in dimensionless form and see them collapse onto a single master curve (Fig. \ref{fig3}(c)). We find that the dimensionless parameter $K_1 n_0 / K_2$, which compares the coagulation rate to the fragmentation rate, captures the evolution of the equilibrium concentration $ n_{\infty}/n_0$ of the system. As $K_1n_0/K_2$ increases, the equilibrium concentration decreases as the system evolves in a more coagulation-dominated regime. In the inset of Fig. \ref{fig3}(c), we report the fragmentation time $r_f^{-1}$   measured across all experiments, $r_f^{-1} = 6.83  \pm $ 1.33~s over the parameter range probed. Comparing  $r_f^{-1}$ to the collision timescale $\tau_c$ (See SI Section III),  we find that a fragmentation event typically occurs once every 10 collisions at early times and once every 6 collisions at later times. We argue that, at early times, a collision is less likely to yield a fragmentation as clusters are smaller. More generally, this trade-off between collision frequency and likelihood of fragmentation qualitatively explains why the overall number of fragmentation events increases linearly with time, as shown in Fig. \ref{fig3}(b). Next, we move to discuss the influence of the packing fraction on the system's dynamics.


\begin{figure*}[!ht]
\begin{center}
    \includegraphics[width=1\textwidth]{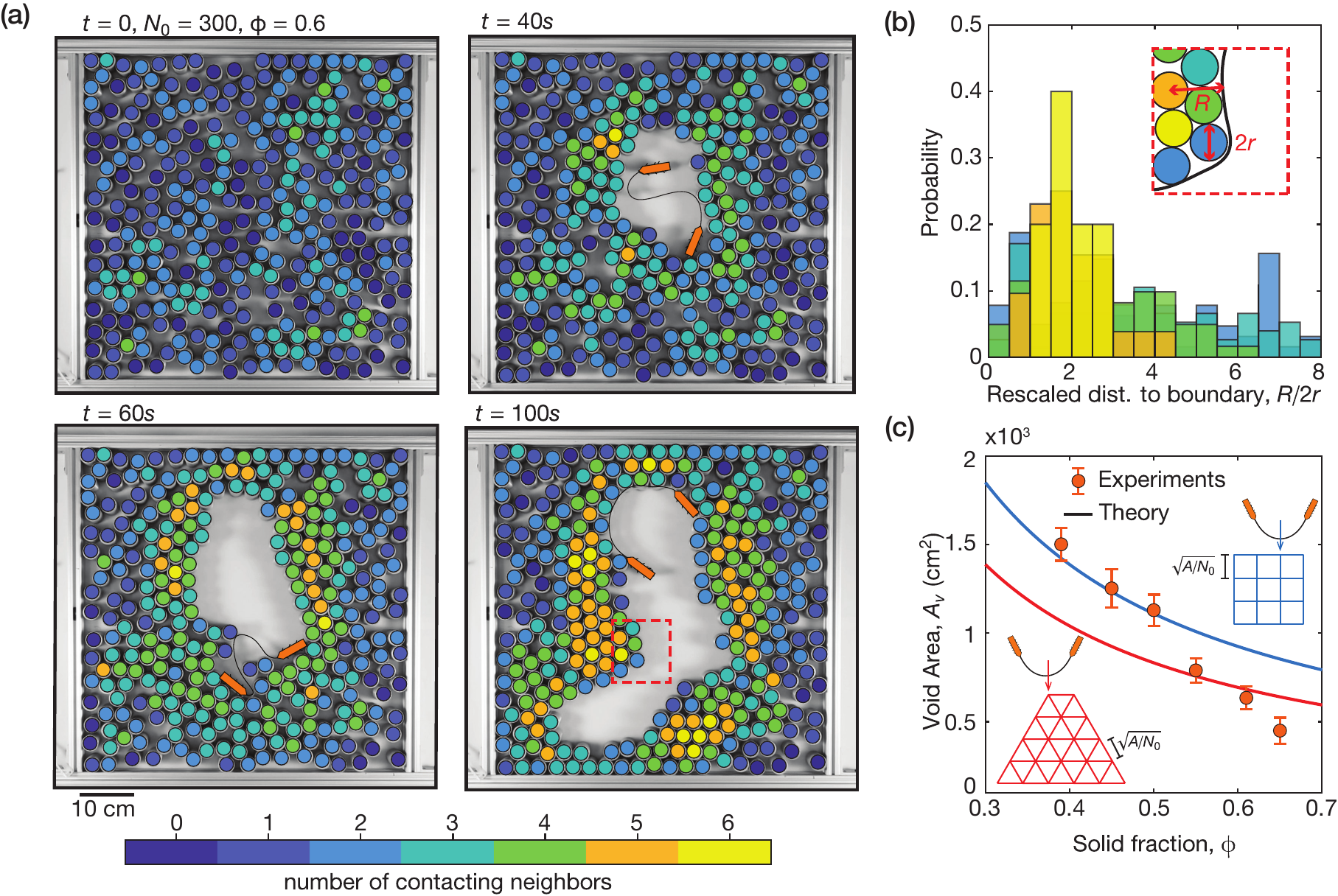}
    \caption{\textbf{Bucklebot navigates in a densely-packed medium} (a) Snapshots of the experiments with densigy $\phi \simeq 0.6$, carved by a bucklebot with rescaled force $F\ell^2/B \simeq 380$. The obstacles are color-coded by the number of contacts they share with their neighbors. (b) Histograms showing the obstacle distance to the boundary $R$ rescaled by their diameter $2r$. They are color-coded by the number of contacting neighbors (same code as (a)). The inset zooms on the area highlighted in (a), defining the Euclidean distance from the obstacle center to the boundary $R$.  (c) We show the void area $A_v$ versus the density $\phi$. The two lines represent two predictions from our geometric model (blue, square lattice; red, hexagonal lattice). Insets: sketch of the corresponding lattices. }
    \label{fig4}
    \end{center}
\end{figure*}

Thus far, we have focused on moderately packed media, defined as granular media where the initial spacing between obstacles is greater than the obstacle size $(\sqrt{A/N_0}-2r)/2r >1$. This limit translates to $\phi <0.2$. However, we observe that in a much denser medium, the bucklebot-obstacle interaction exhibits distinct results. Fig. \ref{fig4}(a) illustrates the scenario where we place a bucklebot in the middle of a dense medium with an area fraction $\phi=0.6$, or $(\sqrt{A/N_0}-2r)/2r=0.14$. The obstacles are initially randomly distributed and color-coded by the number of contacting neighbors in the figure. As our bucklebot displaces surrounding obstacles, we observe the creation of a void surrounded by a boundary with increased packing density (See Movie S2). Over time, the void expands and high-coordination particles accumulate at its boundary. This process continues until the boundary becomes sufficiently thick to resist further displacement, thereby creating a relatively stable enclosure. In Fig. \ref{fig4}(b), we overlay several histograms. They represent the probability of finding elements of a given coordination as a function of their distance to the boundary. We show that obstacles with the most contacting neighbors are predominantly located around 2 obstacle diameters away from the void boundary.  Thus, the boundary that encloses the void is typically 5 obstacles thick. Importantly, this value is consistent with the maximum cluster size the bucklebot can mobilize based on force measurement (See SI Section IV). As such, we have a rationale explaining why voids cease to grow. Using this argument, we can estimate the system's final state. 

In Fig.~\ref{fig4}(c), we report the final void area $A_v$ as a function of the experimental solid fraction $\phi$. We observe that denser systems yield smaller cavities. Additionally, we find that the decay experiences a crisis when $\phi \simeq 0.5$. To rationalize these observations, we quantitatively estimate the void area by modeling it as an effective circular region whose effective radius corresponds to the minimum distance the bucklebot must push outward before a boundary of five obstacle layers forms. To compute this distance, we assume two geometric configurations for the obstacles, as shown in the inset of Fig. \ref{fig4}(c): a square lattice and a hexagonal lattice. In the square lattice case, the minimum distance to access and displace 5 obstacles is $5\sqrt{A/N_0}$. For a hexagonal lattice, the tighter packing reduces this required distance by a factor of $\sqrt{3}/2$.  In Fig. \ref{fig4}(b), we show that, at lower packing fractions, the square lattice assumption provides a reasonable estimate for the void size, as shown by the blue curve. However, as the medium becomes more densely packed and obstacles are forced into more packed arrangements, the hexagonal packing assumption (red curve) yields a better agreement with experimental data. Having established how packing density constrains the extent of void formation, we next examine how bucklebots can leverage local density variations to carve directed paths through dense media.


\begin{figure*}[!ht]
\begin{center}
    \includegraphics[width=.9\textwidth]{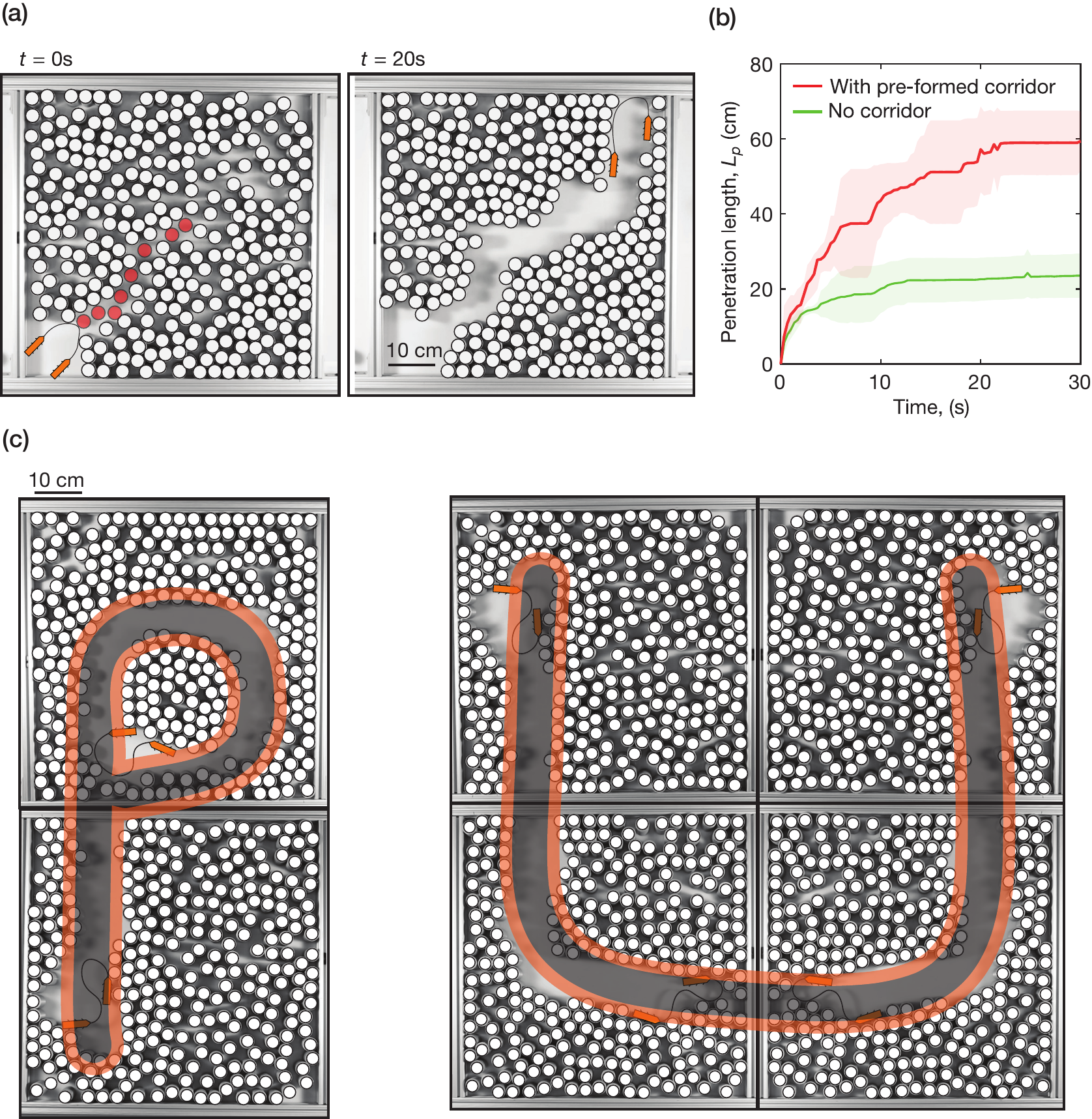}
    \caption{\textbf{Carving and writing in a densely-packed medium} (a) Snapshots of the evolution of the space with $\phi \simeq 0.6$ where a few obstacles (red) are pre-removed and the bucklebot with $F\ell^2/B \simeq 380$ is then sent in. (b) The penetration length $L_p$ is plotted against time for cases with (red) and without (green) pre-removed obstacles. Each shaded area denotes the standard deviation within three trials. (c) Representative trajectories of the bucklebot carving in dense media. Multiple trajectories can be assembled to form the letters "P" and "U". }
    \label{fig5}
    \end{center}
\end{figure*}

Over the course of our experiments in dense media, we observed that our bucklebots preferentially carve along the direction where local packing density is reduced, i.e., the path of least resistance. In Fig. \ref{fig5}(a), we report snapshots of a dense medium  $\phi \simeq 0.6$ in which a few obstacles are selectively removed along a narrow line. We find that the bucklebot follows this path and clears a continuous trajectory along this direction. Fig. \ref{fig5}(b) reports the carving dynamics for this bucklebot by comparing the evolution of the penetration length $L_p$ in the medium with and without such seeds. In all cases, we observe an initial increase in the penetration length $L_p$ before reaching a saturation value. With seeds, the bucklebot reliably detects the low-resistance path and penetrates substantially further. By contrast, in an unmodified medium, $L_p$ saturates at a much shorter distance as the bucklebot easily pushes the obstacles and forms a thick boundary of accumulated obstacles, preventing further advancement (See Video S1). 

This capacity to sense and exploit locally sparse regions highlights the bucklebot's potential for programmable path carving in granular environments. In Fig. \ref{fig5}(c) and Movie S3, we demonstrate that the bucklebot can be guided to extend specific trajectories, which can be assembled to form prescribed figures or shapes. For instance, the letter "P" is assembled by tiling results from two separate trials, while the "U" trajectory combines four experiments. In each trial, we adopt the strategy illustrated in Fig. \ref{fig5}(a): we selectively remove a certain number of obstacles along the intended path to create a subtle low-resistance corridor. The bucklebot is then introduced and penetrates along this corridor, displacing obstacles and leaving behind the desired trajectory. While the examples shown here involve relatively simple shapes, generating more complex patterns will likely require more sophisticated control strategies and larger experimental domains.

To conclude, we have shown that a deformable, elasto-active structure can navigate, restructure, and dynamically reorganize granular media across a wide range of packing fractions. Unlike rigid active systems, Our bucklebot leverages its compliance and morphology to achieve multiple behaviors such as clustering, void carving, and directional path-following without explicit external guidance. Our combined experimental and theoretical approaches reveal how local obstacle density and force transmission govern these emergent behaviors. This understanding enables programmable environmental reconfiguration, including tasks such as selective aggregation, persistent void creation, and the formation of programmable trajectories. We note that our experiments are conducted in controlled laboratory conditions with relatively simple granular configurations. Extending these concepts to more complex and larger environments, such as heterogeneous granular configurations or frictional terrains, will be the topic of further work. Additionally, while we have focused primarily on a single elasto-active structure operating in a beam-dominated regime, exploring systems where multiple bucklebots cooperate, or where bucklebots interact with purely active agents, such as individual microbots, may reveal rich collective behaviors and new modes of environmental transformation.

\section*{Materials and Methods}
The active microbots are commercially available battery-powered vibrating microbots (Hexbug Nano). Each microbot has a length of 45 mm, a width of 15 mm, a height of 15 mm, and a mass of 7.5 g. Its motion is generated from an internal vibration of a rotating motor transmitted to 12 soft rubber legs to achieve a speed of approximately $154 \pm15$ mm/s. The elastic beams are fabricated by laser-cutting polyester shim stock (Epilog Helix-60 Laser Engraver) with an elastic modulus of 2.9 GPa. The thickness and length of these beams are calibrated to produce variations in bending stiffness across experiments. The collars connecting the microbots and elastic beams are designed in Rhino and 3D-printed on a Prusa i3 printer using polylactic acid (PLA)(density $\rho$ = 1.2 g/cm\textsuperscript{3} and elastic modulus $E$ = 5 GPa). The beams are clamped to the collars using Dodge 0-80 .115 inch length inserts and corresponding screws. 

The active force exerted by the bucklebot is estimated via an Instron 10N load cell. Two characteristic forces are measured: the front impact force, obtained by sending the bucklebot vertically towards the load cell, and the flapping force, which arises from the lateral oscillations as the bucklebot slides along the cell due to motor-induced vibrations. The cylindrical obstacles used in experiments are 3D printed using polylactic acid (PLA) filament. Each obstacle weighs approximately $2$ g. For visual tracking, a square red tape marker is affixed to the top surface of each obstacle. To capture the evolution of the obstacles and clusters, a Canon EOS 80D camera is held by a frame looking down at a large white cast acrylic sheet from McMaster-Carr on top of the lab table.

\section*{Acknowledgements}
This work was supported by NSF Grant CMMI 2343539 and NSF Future Manufacturing Grant CMMI 2037097.

\bibliographystyle{unsrt}  
\bibliography{lit.bib}  

\newpage

\end{document}